\documentclass[twocolumn]{aastex631}

\begin{document}

%\title{Template \aastex Article with Examples: 
%v6.3.1\footnote{Released on March, 1st, 2021}}
\title{Dust hot spots at 10 au scales around the Class 0 binary IRAS 16293-2422 A: \\ a departure from the passive irradiation model}

\correspondingauthor{Mar\'ia Jos\'e Maureira}
\email{maureira@mpe.mpg.de}

\author[0000-0002-7026-8163]{Mar\'ia Jos\'e Maureira}
\affiliation{Max-Planck-Institut für extraterrestrische Physik (MPE), Gießenbachstr. 1, D-85741 Garching, Germany}

\author[0000-0003-1613-6263]{Munan Gong}
\affiliation{Max-Planck-Institut für extraterrestrische Physik (MPE), Gießenbachstr. 1, D-85741 Garching, Germany}

\author[0000-0002-3972-1978]{Jaime E. Pineda}
\affiliation{Max-Planck-Institut für extraterrestrische Physik (MPE), Gießenbachstr. 1, D-85741 Garching, Germany}

\author[0000-0003-2300-2626]{Hauyu Baobab Liu}
\affiliation{Academia Sinica Institute of Astronomy and Astrophysics, P.O. Box 23-141, Taipei 10617, Taiwan}

\author[0000-0003-1572-0505]{Kedron Silsbee}
\affiliation{Max-Planck-Institut für extraterrestrische Physik (MPE), Gießenbachstr. 1, D-85741 Garching, Germany}
\affiliation{The University of Texas at El Paso, 500 W University Ave, El Paso, TX 79968, USA}

\author[0000-0003-1481-7911]{Paola Caselli}
\affiliation{Max-Planck-Institut für extraterrestrische Physik (MPE), Gießenbachstr. 1, D-85741 Garching, Germany}

\author[0000-0002-2887-2287]{Joaquin Zamponi}
\affiliation{Max-Planck-Institut für extraterrestrische Physik (MPE), Gießenbachstr. 1, D-85741 Garching, Germany}

\author[0000-0003-3172-6763]{Dominique M. Segura-Cox}
\affiliation{Department of Astronomy, The University of Texas at Austin, 2515 Speedway, Austin, TX 78712, USA}

\author[0000-0002-1730-8832]{Anika Schmiedeke}
\affiliation{Green Bank Observatory, PO Box 2, Green Bank, WV 24944, USA}

% \author{August Muench}
% \affiliation{American Astronomical Society \\
% 1667 K Street NW, Suite 800 \\
% Washington, DC 20006, USA}

% %\collaboration{20}{(AAS Journals Data Editors)}

% \author{F.X Timmes}
% \affiliation{Arizona State University}
% \affiliation{AAS Journals Associate Editor-in-Chief}

% \author{Amy Hendrickson}
% \altaffiliation{AASTeX v6+ programmer}
% \affiliation{TeXnology Inc.}

% \author{Julie Steffen}
% \affiliation{AAS Director of Publishing}
% \affiliation{American Astronomical Society \\
% 1667 K Street NW, Suite 800 \\
% Washington, DC 20006, USA}



\begin{abstract}

Characterizing the physical conditions at disk scales in Class 0 sources is crucial for constraining the protostellar accretion process and the initial conditions for planet formation. We use ALMA 1.3 mm and 3 mm observations to investigate the physical conditions of the dust around the Class 0 binary IRAS 16293-2422 A (sep $<$100 au) down to $\sim$10 au scales. The circumbinary material's spectral index, $\alpha$, has a median of 3.1 and a dispersion of $\sim$0.2, providing no firm evidence of mm-sizes grains therein. Continuum substructures with brightness temperature peaks of $T_{\rm b}\sim$60-80 K at 1.3 mm are observed near the disks at both wavelengths. These peaks do not overlap with strong variations of $\alpha$, indicating they trace high-temperature spots instead of regions with significant optical depth variations. The lower limits to the inferred dust temperature in the hot spots are 122, 87 and 49 K. Depending on the assumed dust opacity index, these values can be several times higher. They overlap with high gas temperatures and enhanced complex organic molecular (COM) emission. This newly resolved
dust temperature distribution is in better agreement with the expectations from mechanical instead of the most commonly assumed radiative heating. In particular, we find that the temperatures agree with shock heating predictions. This evidence and recent studies highlighting accretion heating in Class 0 disks suggest that mechanical heating (shocks, dissipation powered by accretion, etc.) is important during the early stages and should be considered when modeling and measuring properties of deeply embedded protostars and disks.

\end{abstract}



\section{Introduction} 
\label{sec:intro}

Since disk structures around the more evolved Class II sources can be observed directly, i.e., without much confusion from an infalling envelope, they have been studied in more  detail than their younger and embedded Class 0 and I counterparts. A model of disk density and thermal structure is well established \citep{1997ChiangSpectral,2007DullemondModels,2013HenningChemistry,2021ObergAstrochemistry}. Stellar irradiation is commonly the dominant process setting the temperature in Class II/III disks, and hence they are described as passive disks. This passively irradiated disk model results in a shielded and dense midplane with cold ($\lesssim 20$ K) temperatures, a warm surface layer, and a hot ($\gtrsim100$ K) low-density atmosphere \citep{2013HenningChemistry}. Such temperature structure is supported by observations \citep{2018PinteDirect,2020rabInterpreting}. Accretion heating is important in the inner parts of T Tauri protoplanetary disks, while at large radii irradiation dominates. Assuming the accretion rate of typical T Tauri stars, viscous heating would only be significant in the inner $\sim$10 au in these disks \citep{2007DullemondModels}. \\

Whether this passive heating model is valid for the disk and inner envelope of early stage protostellar sources is not yet clear (e.g., \citealt{2021LiuMagnetically,2021ZamponiYoung,2021XuFormation}). In embedded and actively accreting sources, mechanical heating (due to shocks, viscous dissipation, adiabatic compression, etc.) can be a crucial ingredient in modelling the temperature of both gas and dust at disk scales due to the higher densities and accretion rates at these stages. Considering such non-radiative heating mechanisms can have important implications for the estimations of embedded disk masses \citep{2018GalvanMadridEffects,2021ZamponiYoung,2022XuTesting}, envelope and disk grain sizes \citep{2017LiSpectralIndex,2021ZamponiYoung}, and for the disk chemistry \citep{2011IleeChemistry,2014NaturSakai,2015HarsonoVolatile,2015EvansGravitational,2017IleeFragments,2019EvansMolecular,2020Belloche2020Questioning, 2021ZamponiYoung, 2022VastelHotMethanol}. \\

In particular, shock and adiabatic compression heating can be present in regions where the infalling material is deposited into a disk \citep{2019ArturdelaVillarmoisPhysical,2014SakaiChange,2017MiuraComprehensive,2022PinedaPPVII}, near spiral arms in a gravitationally unstable disk \citep{2011IleeChemistry,2008BoleyGravitationalInstabilities,2001BossDiskInstability,2007BossTestingDiskInstability} or in the material around compact binaries \citep{2019MostaGas,2022VastelHotMethanol}. This type of heating can lead to local temperature enhancements and thus can be more easily decoupled from other types of heating such as irradiation or steady-state viscous heating for which a smooth decrease of temperature with increasing distance to the protostar(s) is expected. \\

In this study, we find that the continuum substructures around the circumstellar disks in a Class 0 binary correspond to localized dust temperature enhancements. The two protostars (IRAS 16293-2422 A1-A2) are separated by a projected distance of 54 au ($d=141$ pc, \citealt{2018OrtizLeonGaia}). The binary is also separated by $\sim$700 au from the single protostar B, forming together a hierarchical triple system (Figure~\ref{fig:1mm_3mm_alpha}).  \cite{2020OyaSubstructures} studied the gas temperature around A1-A2 using high spatial resolution observations ($\sim$14 au) and measured temperature peaks of 300 K or higher, offset by 20-30 au from the nearest protostar. Here, we demonstrate that both the gas and dust temperatures have similar spatial distributions consistent with localized hot spots in agreement with mechanical instead of radiative heating.

The paper is organized as follows: in Section~\ref{sec:obs} we describe the observations and data reduction. In Section~\ref{sec:results} we present the observations and spectral index map. In Section~\ref{sec:analysis} we derive the dust temperature, optical depth variations, constraints for the dust opacity index and comparison with the region with enhanced Complex Organic Molecular (COM) emission. Section~\ref{sec:discussion} corresponds to the discussion and Section~\ref{sec:conclusions} to the summary and conclusions. 

\begin{figure*}[t!]
   \centering
     \includegraphics[width=0.95\textwidth]{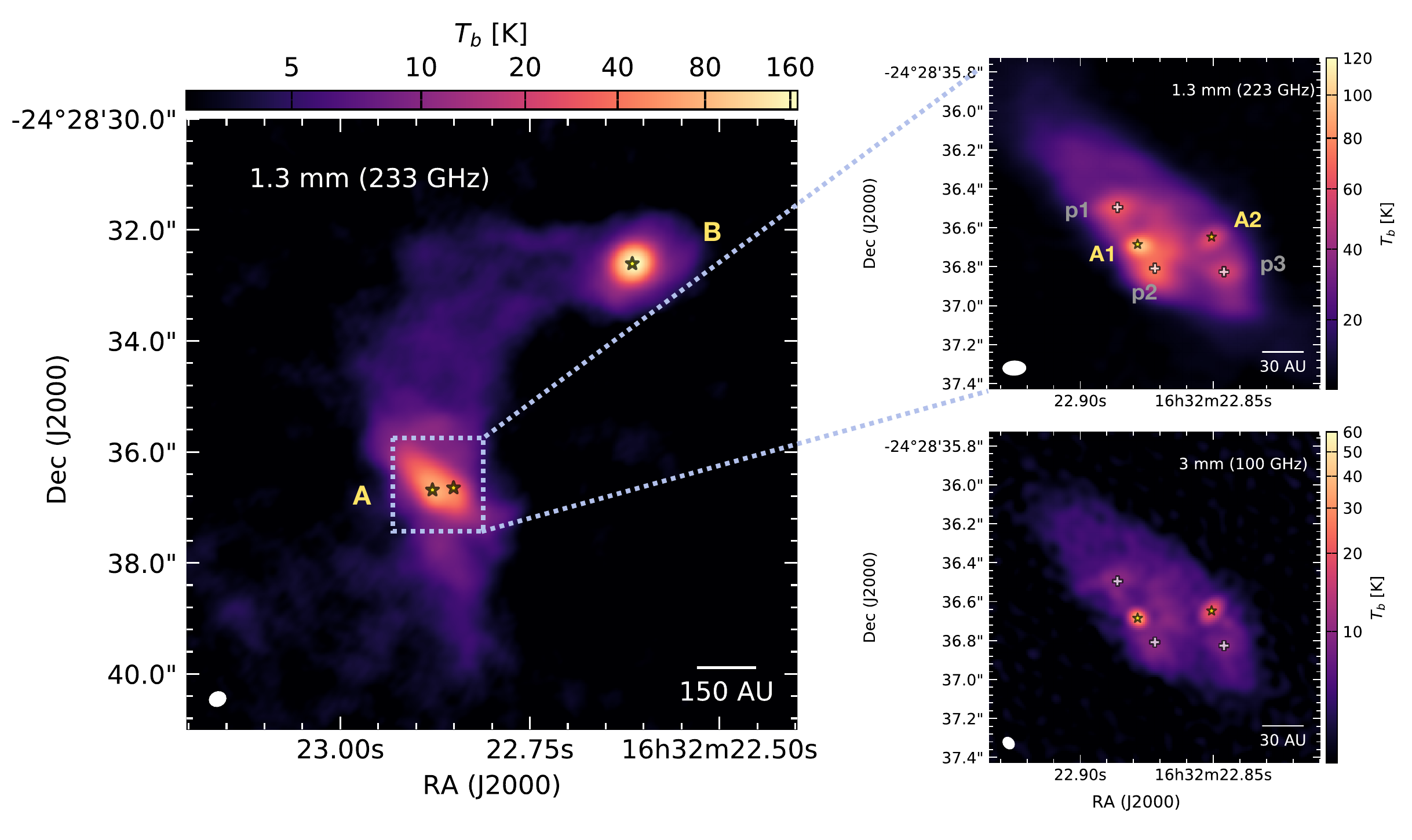}
      \caption{ {\it Left:} ALMA 1.3 mm observations towards the triple Class 0 system IRAS 16293-2422. The observations have a resolution of $\sim$35 au and were published in \cite{2018SadavoyDustIRAS16293}. {\it Right:} ALMA 1.3 mm and 3 mm observations of the southern binary system at a resolution of $\sim$13 au and $\sim7$ au respectively. Similar maps for source B are published in \cite{2021ZamponiYoung}. In all panels the position of the protostars are marked with a star symbol. Additional continuum peaks around the A1 and A2 protostars are labeled and marked with a cross in the right panels. The beam is shown in the bottom left corner of each panel. }
         \label{fig:obs_1mm_3mm}
\end{figure*}

\section{Data}
\label{sec:obs}

The ALMA band 6 observations of IRAS 16293-2422 were taken during 2017 August 21. The observations were part of the cycle 4 project ID:2016.1.00457.S (PI: Y. Oya). We used the extended configuration data available in this project which is best suited for comparison with the band 3 data (see below). It consists of baselines ranging between 21 and 3697 m, with a maximum recoverable scale of $\sim$0.6" (or 85 au). The pointing center of the observations was ICRS 16:32:22.87 -24:28:36.5. The bandpass/flux calibrator and phase calibrator were J1517-2422 and J1625-2527, respectively, the same as for the band 3 observations. The spectral setup consisted of four spectral windows centered at frequencies of 240.3 GHz, 240.5 GHz, 224.7 GHz, 222.9 GHz, with a channel width of 61.0 kHz, 15.3 kHz, 30.5 kHz, 488.3 kHz and bandwidth of 0.23 GHz, 0.06 GHz,  0.11 GHz, 0.94 GHz, respectively. We use CASA 5.6.2 \citep{2007McMullinCASA} to calibrate and image the data. Calibration of the raw visibility data was done using the standard pipeline. To create the continuum image we carefully check each spectral window and flag lines. The resultant bandwidth for the band 6 continuum is 0.12 GHz. When imaging the continuum, we iteratively performed phase-only self-calibration reaching a minimum solution interval of solint$=$\textquotesingle int\textquotesingle. Afterwards, we performed one amplitude self-calibration, with solint$=$\textquotesingle inf\textquotesingle\ and combine$=$\textquotesingle scan,spw\textquotesingle\ (one solution per antenna per track). For this last step we used calmode$=$\textquotesingle a\textquotesingle\ and provided the phase-only solutions as gaintable to be applied on the fly. For the process of self-calibration we selected visibilities with a minimum baseline of 120 k$\lambda$, in order to avoid missing flux artifacts, but also because shorter baselines are not covered by the band 3 observations with which we want to compare for our analysis. \\

The ALMA band 3 observations  of IRAS 16293-2422 (ID:2017.1.01247.S, PI: G. Dipierro) were taken during 2017 October 8 and 12 using the most extended Cycle 5 configuration of ALMA (41.4 m–16.2 km baseline range) and a maximum recoverable scale of $~\sim$0.5" (or 70 au). The pointing center of the observations was ICRS 16:32:22.63 -24:28:31.8. The bandpass/flux calibrator and phase calibrator were J1517-2422 and J1625-2527, respectively. The single spectral window used in this work consists of 128 channels centered at 99.988 GHz, with a total bandwidth of 2 GHz. The procedure for the calibration and imaging of the continuum data (including phase and amplitude self-calibration) are detailed in \cite{2020MaureiraOrbital}. To address whether the resultant continuum image was significantly affected by line contamination, we also created an image using the line-free channels in the other four narrow windows in the setup. These windows consisted of 960, 1920, 960 and 1920 channels with widths of 22.070 kHz and centered at the frequencies of $^{13}$CO (1-0), C$^{17}$O (1-0),  C$^{18}$O (1-0), and CS (2-1), respectively. The observed morphologies were consistent between the two continuum images and the intensities were in good agreement as well.\\

%The final continuum dataset after phase+amplitude self-calibration was imaged using the tclean task with multiscale and a robust parameter of 0.5. {\bf The beam size, beam P.A. and noise of the final continuum image are 0.048"$\times$0.044" (6.5 au), 79.3$^{\circ}$ and 15 $\mu$Jy beam$^{-1}$, respectively. 

The continuum images for both bands (after phase+amplitude self-calibration) were created in CASA 5.6.2 using the task tclean. In both cases the imaging was done with the multiscale deconvolver using 3 scales (point source, beam, and 3$\times$beam), with a robust parameter of 0. Further, in both cases we use a uv-range parameter of 120-2670 k$\lambda$, corresponding to the overlapping baselines between the datasets. A reference frequency of 223 GHz and 100 GHz was set for the band 6 and band 3 observations, respectively. The resultant beam size, P.A. and rms of the band 6 observations are 0.114"$\times$0.069", -88.2$^{\circ}$ and 104 $\mu$Jy beam$^{-1}$, respectively. For the band 3 observations these parameters correspond to 0.062"$\times$0.050", 41.7$^{\circ}$ and 17 $\mu$Jy beam$^{-1}$, respectively. For the band 3 we did an additional image for the purpose of the spectral index map. To match the resolution between the two bands, we applied a Gaussian taper (uvtaper$=[0.13",0.07",110^\circ$]) in tclean to the band 3 observations so the resultant beam was as similar as possible to the beam at 1.3 mm. We then applied a restoring beam equal to the 1.3mm beam. The resultant rms of the matching beams band 3 observations is 24 $\mu$Jy beam$^{-1}$. This study focuses on source A in IRAS16293-2422 (hereafter IRAS 16293 A), corresponding to the binary pair (A1-A2) within the triple system (Figure~\ref{fig:obs_1mm_3mm}). The corresponding maps for source B, are presented in \cite{2021ZamponiYoung}.

\section{Results}
\label{sec:results}

\subsection{Morphological comparison between wavelengths}

Figure~\ref{fig:obs_1mm_3mm} shows the continuum images at 1.3 mm and 3 mm towards IRAS 16293 A. The protostellar disks towards A1 and A2 are clearly distinguished, corresponding to the only bright and compact sources in the 3 mm map. The circumstellar disks are surrounded by a larger disk-like structure with a semi-major axis of about 110 au. Previous studies show that rotation dominates the kinematics of this structure. However, the velocity as a function of distance cannot be uniquely determined as Keplerian. Infall and rotation with conservation of angular momentum can also explain the velocity pattern \citep{2012PinedaFirst,2014FavreDynamical,2020MaureiraOrbital,2020OyaSubstructures}. Because of this we refer generically to this structure as circumbinary material. 

In the circumbinary material, additional seemingly compact structures (p1, p2 and p3) are more clearly observed in the 1.3 mm map\footnote{p1 and p2 closely correspond to the submm peaks labeled Aa and Ab in \cite{2005ChandlerIRAS16293}.}. \cite{2020OyaSubstructures} claimed that these peaks possibly reveal the location of additional protostellar sources. However, Figure~\ref{fig:obs_1mm_3mm} shows that p1, p2 and p3 have clear counterparts at 3 mm, previously seen in \cite{2020MaureiraOrbital}. Unlike the circumstellar disks, the peaks p1, p2 and p3 do not appear compact at 3 mm. We note that at 1.3 mm, p1 and p2 have even higher or comparable brightness temperatures ($T_{\rm b}\approx57-76$ K) than the A2 disk ($T_{\rm b}\approx62$ K). This can certainly lead to confusion about the true multiplicity of the system, as well as the true location of the prototellar sources in the 1.3 mm observations when other (longer wavelengths) observations are not available. It is only in the 3 mm observations, when the optical depth decreases, that one can unambiguously confirm the binary nature of the system (at least down to the current best resolution of 6.5 au at 3 mm).

\subsection{Spectral index map}
\label{sec:spectral_index}

We calculated the spectral index $\alpha$ using the 1.3 mm and 3 mm maps imaged with the same {\it uv}-range. To address the beam differences, we first cleaned the 3 mm data applying a Gaussian taper to closely match the 1.3 mm beam and use a restoring beam equal to the beam at 1.3 mm (Section~\ref{sec:obs}). The spectral index map was created using
\begin{equation}
\alpha = \frac{\ln(I_{\nu_1}/I_{\nu_2})}{\ln(\nu_1/\nu_2)}, % \nonumber
\label{eq:spec_index}
\end{equation}
where $\nu_1$ and $\nu_2$ correspond to 223 GHz and 100 GHz, respectively. To estimate the error in each pixel we first calculated the propagated errors using the rms of each image. We then added these values in quadrature to the flux calibration error in $\alpha$. The 1$\sigma$ flux calibration error corresponds to 2.5\% and 5\% for band 3 and 6 respectively\footnote{See Cycle 9 ALMA Technical handbook, Section 10.2.6.} resulting in an absolute calibration error in $\alpha$ of 0.07. In order to keep the uncertainty low when studying variations across the region, we excluded all pixels which have a combined total error higher than 0.13. Thus, the resultant $\alpha$ map has a systematic 1$\sigma$ error of 0.07, and variations from point to point have a $1\sigma$ error $\leq0.1$. \\

\begin{figure}[ht!]
   \centering
     \includegraphics[width=0.5\textwidth]{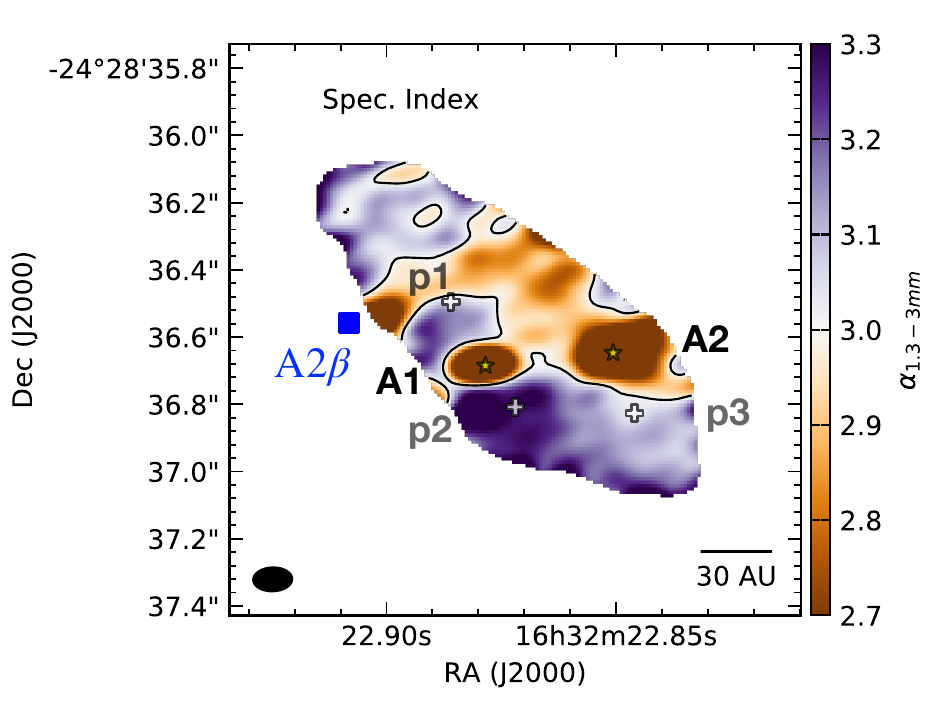}
      \caption{1.3-3 mm spectral index towards the IRAS 16293 A binary. Black contours on the spectral index panels are drawn for $\alpha=$ 3.0. Star and cross markers show the protostars and the additional continuum peaks similar to Figure~\ref{fig:obs_1mm_3mm}. The free-free ejecta A2$\beta$ estimated location is marked with a blue square, using the proper motions from \cite{2019HernandezGomezNature}. The beam is shown in the bottom left corner.
                  }
         \label{fig:1mm_3mm_alpha}
\end{figure}

Figure~\ref{fig:1mm_3mm_alpha} shows the resultant 1.3-3 mm spectral index map. In the circumbinary material, the spatial distribution of $\alpha$ does not show any clear axisymmetric symmetry or monotonic behaviour with respect to the center of the circumbinary material or protostellar disks. To quantify the distribution of $\alpha$ values, we calculated the median and dispersion around the median with a Kernel Density Estimation (KDE). For this calculation we masked out the circumstellar disks. Figure~\ref{fig:sourceA_alpha_kde} in the Appendix shows the distribution of the spectral index and the pixels considered for the calculation. The resultant median is $\alpha=3.1$. The value of $\alpha$ is $\gtrsim2.9$ in most places ($84\%$) with a dispersion of less than 10\% around the median.\\

The highest values of $\alpha$ in the map (up to $\sim$3.3) are observed around the A1 circumstellar disk, at the location of p2 and near p1. Lower values in the circumbinary material are enclosed by the $\alpha=3$ contour which, besides the circumstellar disks, is enclosing regions to the north and northeast of A1 and A2, respectively. In these regions the value of $\alpha$ is $\sim$2.9. To the northeast of A1 and close to the border of the circumbinary structure, there is a spot reaching $\alpha=2.4$. The low values therein could be simply to spatial scale sensitivity and beam convolution but if real, it could be due the presence of nearby free-free emission. In particular, there is known ejecta whose origin is thought to be the A2 protostar \citep{2010PechConfirmation,2019HernandezGomezNature}. The blue cross marks the inferred position of this nearby clump called A2$\beta$, estimated using the proper motions reported in \cite{2019HernandezGomezNature}. 

\section{Analysis}
\label{sec:analysis}

%\subsection{Relation between $\alpha$ and $\tau_{\nu}$}
%\label{sec:tau_alpha}
Spectral index variations can be used to investigate the nature of the observed continuum substructures (p1, p2 and p3) and the dust properties in the circumbinary material. This is because of the relation between $\alpha$ and the optical depth $\tau_{\nu}$. The relation can be understood by approximating the continuum emission with a modified black body \citep{1983HIldebrandDetermination}:

\begin{equation}
     I_{\nu}=B_{\nu}(T)(1-e^{-\tau_{\nu}}),
     \label{eq:modi_bb}
\end{equation}

where $T$ is the dust temperature, $B_{\nu}(T)$ is the Planck function at temperature $T$ and $\nu$ is the frequency. If the emission at two or more frequencies traces the same material at a given temperature $T$ and $B_{\nu}(T)\propto \nu^2T$ (Rayleigh-Jeans regime), $\alpha$ only depends on the optical depth and its variations with frequency. For thermal dust emission the optical depth can be described as $\tau_{\nu}=\kappa_\nu\Sigma$ where $\Sigma$ is the dust surface density and $\kappa_\nu$ is the dust opacity. The latter can be described as $\kappa_\nu$=$\kappa_1(\nu/\nu_1)^{\beta}$ where $\nu_1$ is a reference frequency (in our case $\nu_1=223$ GHz) and $\beta$ is the dust opacity index. 

The above assumptions lead to a range of $\alpha$ values limited to $2\leqslant\alpha\leqslant2+ \beta$ with the lower and upper bounds corresponding to the optically thick and optically thin limits, respectively. Thus, if the emission is fully optically thin, $\alpha$ and $\beta$ changes are related as $\alpha=2+\beta$. In this case, $\alpha$ variations fully trace dust property changes such as grain sizes \citep{2014TestiDustEvolution}. However, if the emission is not fully optically thin, lower values of $\alpha$ (at a fixed $\beta$) are related to an increase of $\Sigma$ and thus physical properties of the structure.

\subsection{Spectral index in circumbinary material}%emissivity index}
\label{sec:alpha_var}

As discussed in Section~\ref{sec:spectral_index}, $\alpha$ in the circumbinary material is $\geqslant$2.9 in most places and the dispersion around the median value of 3.1 is below 10\%. The small $\alpha$ variations together with the observed spatial distribution in Figure~\ref{fig:1mm_3mm_alpha} favor a scenario in which $\beta$ remains constant throughout the structure. This is because if the small $\alpha$ variations were due to changes in $\beta$ (as $\beta=\alpha-2$) it would imply that the grain sizes were smallest near the circumstellar disks and larger further away, which would be surprising. In particular, the grain sizes would be the smallest ($\beta=1.3$) around A1, with increasing sizes farther away from the disk ($\beta$=0.7-0.9 at around 30-40 au from the circumstellar disk center) and intermediate sizes ($\beta$=1-1.1) up to 100 au from both disks (North-East corner). Moreover, this scenario would imply that the emission is optically thin, and the low optical depths at 1.3 mm together with the observed high brightness temperature at p1 and p2 would lead to dust temperatures well above 500 K  (assuming $\tau\lesssim0.1$) at 30-40 au (projected) from the circumstellar disks.\\

Assuming that $\beta$ remains constant throughout the observed circumbinary structure, the narrow range of variations in Figure~\ref{fig:1mm_3mm_alpha} can be related to changes in the physical properties of the circumbinary material such as surface density. In this case the surface density in the circumbinary material has to be high enough to depart from the fully optically thin regime. For optical depths between 0.2 and 1 at 223 GHz, and assuming a conservative dust opacity range of 0.9-2 cm$^{2}$ g$^{-1}$, we find dust surface densities of 0.1-1.1  g cm$^{-2}$.  Assuming a gas to dust ratio of 100, these values are in good agreement with recent simulations of the formation of protostellar binaries from the collapse of an initial 3.7 M$_{\odot}$ core \citep{2020SaikiTwin}, comparable with the amount of material surrounding the triple system measured with Herschel observations \citep{2020LadjelateHerschel}. In this scenario in which $\beta$ is constant, our observations provide a lower limit to $\beta$. Since the values of $\alpha$ can only vary between $2\leqslant\alpha\leqslant2+ \beta$, the full range of $\alpha$ values observed in the circumbinary structure implies a conservative lower limit to $\beta$ of 1.3 ($\alpha=3.3$, to recover the values near p2, see Figure~\ref{fig:sourceA_profiles}). Higher values such as those consistent with the ISM ($\beta\sim$1.7) would also be in agreement with the observations, and would imply higher optical depths than those obtained assuming a $\beta$ value of 1.3.

\subsection{Spectral index and continuum substructures}
\label{sec:alpha_substructures}

\begin{figure*}
   \centering
     \includegraphics[width=1\textwidth]{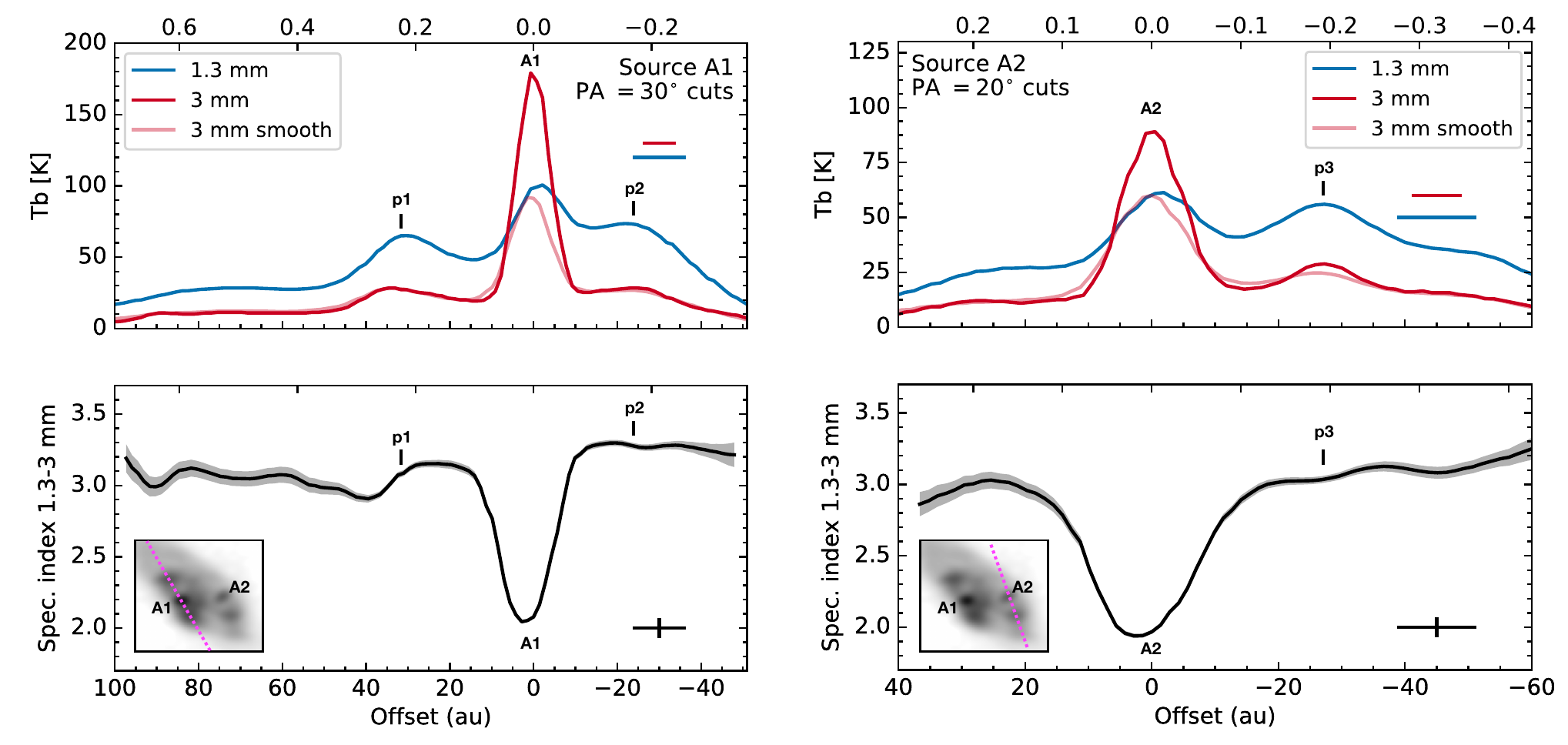}
      \caption{Source A intensity profiles at 1.3 mm and 3 mm and spectral index. The directions of the cuts for the profiles are shown in insets within the bottom panels. Left: intensity and spectral index profiles centered at A1 and with P.A$=30^{\circ}$. Right: Intensity and spectral index profiles centered at A2 and with P.A$=20^{\circ}$. The shaded region around the spectral index curve indicates only the statistical uncertainties. There is also a systematic uncertainty of 0.07 from the flux calibration, corresponding to half the size of the vertical line shown on the corner of the bottom panels. The beam size for the spectral index value is shown with a horizontal line. In all panels, the bottom x-axis is in units of au while the top x-axis in units of arcseconds.}
         \label{fig:sourceA_profiles}
\end{figure*}

Figure~\ref{fig:sourceA_profiles} shows brightness temperature and spectral index profiles along cuts centered on the protostars and passing through the continuum substructures around them. The two circumstellar disks can be clearly identified as the location where the $T_{\rm b}$ at 3 mm is the highest, corresponding to the sharp decrease on $\alpha$, reaching values close to 2 i.e., the optically thick limit. We note that the resultant low spectral index values for the circumstellar disks have to be interpreted with caution when analyzing the thermal dust emission. This is because at 3 mm the emission can be contaminated by free-free \citep{2019HernandezGomezNature,2020MaureiraOrbital}. The focus of this work is on the material outside the circumstellar disks, but to simply illustrate the uncertainty in the circumstellar disks, the values of $\alpha$ can increase to 2.1-2.6 at both circumstellar disk's center if free-free contamination is $10-40\%$, respectively.\\

The peaks p1, p2 and p3 are not always directly associated with a decrease in $\alpha$ as in the case of the circumstellar disks. The clearest example corresponds to the case of p2 (offset $\sim-30$ au in the left panels of Figure~\ref{fig:sourceA_profiles}). This $T_{\rm b}$ peak does not correspond to a significant decrease in $\alpha$, which shows instead a flat profile with a value of $\sim3.3$ and variations of 1\%. The $T_{\rm b}$ peak p1 (offset $\sim$30 au in the left panels) is first associated with a flat $\alpha\sim3.2$ profile closer to the disk, then followed by a decrease in $\alpha$ down to $\sim2.9$. These lower values correspond to the orange color region enclosed by contours in Figure~\ref{fig:1mm_3mm_alpha}. Lastly, the $T_{\rm b}$ peak p3 (offset $\sim$-30 au in the right panels) corresponds with smooth $\alpha$ values (less than $5\%$ variations). These behaviors suggest that p1, p2 and p3 cannot be fully explained by changes in the optical depth alone, otherwise we would see correlated $\alpha$ changes. It follows that an increase of dust temperature at the location of these peaks is required to explain both the rather constant $\alpha$ values and the increase in the observed flux therein.\\

\subsection{Temperature and $\tau$ maps}
\label{sec:temp_tau_derivation}

\begin{figure*}
   \centering
     \includegraphics[width=\hsize]{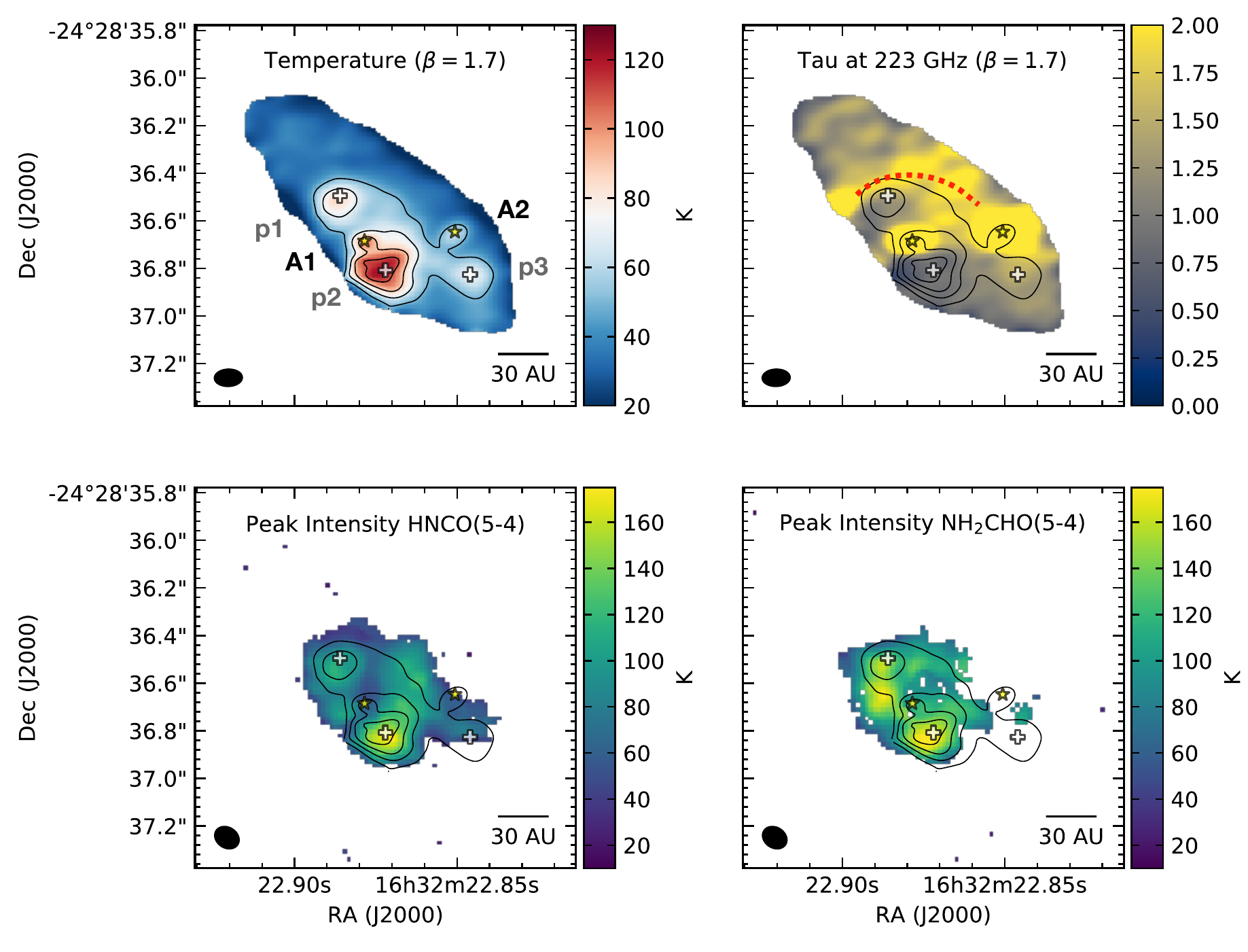}
      \caption{Top: Derived dust temperature and optical depth at 1.3 mm (223 GHz) maps assuming a constant dust opacity index of $\beta=1.7$ (i.e., ISM dust). Bottom: Line peak intensity of HNCO(5-4) and NH$_2$CHO(5-4) from a Gaussian fit of the data presented in \citep{2020MaureiraOrbital}. Black contours in all panels follow the dust temperature (top left panel) and are drawn at 50, 70, 90 and 110 K. Star and cross markers show the protostars and the additional continuum peaks similar to Figure~\ref{fig:obs_1mm_3mm}. The red dotted line highlights the region where the optical depth values at 1.3mm are the highest outside the circumstellar disks. The beam corresponding to each color map is shown in the bottom left corner. }
         \label{fig:tau_temp_beta_ism}
\end{figure*}

Following the discussion in Section~\ref{sec:alpha_var}, we derived dust temperature and $\tau_{\rm 1.3 mm}$ maps assuming a constant $\beta$ value. Dust temperature and $\tau_{\rm 1.3 mm}$ were computed for each pixel based on the spectral index and 1.3 mm continuum maps, assuming modified black body emission (Equation~\ref{eq:modi_bb}) and full Planck function for $B_{\nu}(T)$. We note that the Rayleigh–Jeans approximation gives similar results as it remains valid in most of the map. \\

As a conservative scenario, in Figure~\ref{fig:tau_temp_beta_ism} we show the results for a constant $\beta$ value of 1.7 (consistent with ISM dust) from which we get an upper limit of absolute optical depth values and a lower limit of the absolute dust temperature values. The spatial temperature and optical depth variations are robust regardless of the fixed $\beta$ used. Local temperature enhancements away from the circumstellar disks are more clearly related to the 1.3 mm T$_b$ peaks p1, p2 and p3. These local hot spots outside the disks show temperature peaks of 122 K (p2), 87 K (p1) and 49 K (p3), and these values are only lower limits. The calculated dust temperature increases by about a factor of 2 and 3 for p1 and p3, respectively, if $\beta=1.3$ is assumed instead (see Appendix Figure~\ref{fig:temp_tau_1.3}). In the case of p2 the optical depth and temperature are degenerate when assuming $\beta=1.3$ (optically thin regime). Assuming $\tau_{1.3mm}=0.3$ at p2 results in a dust temperature of $\sim$250 K. Thus the temperature at p2 could be at least a factor of 2 higher than our derived lower limit in Figure~\ref{fig:tau_temp_beta_ism} using $\beta=1.7$.\\

These local hot dust spots correlate very well with the localized gas temperature enhancements derived in \cite{2020OyaSubstructures} at 10 au scales using integrated intensity ratios between two H$_2$CS transitions. Their derived gas temperature peaks for p2 and p1 are at about 300 K and 200 K, respectively, while gas temperatures between 200 and 300 K are observed towards p3. These values are 2-5$\times$ the dust temperature we derived assuming $\beta=1.7$, while the gas temperature are only 1-2$\times$ the dust temperatures when assuming $\beta=1.3$. Overall, our dust hot spots have temperatures equal to or lower than the corresponding gas temperatures in \cite{2020OyaSubstructures}.  \\

The optical depth map shows a region to the North of the disks, with values up to $\times2$ higher than the surroundings (see red dotted line in Figure~\ref{fig:tau_temp_beta_ism}). If due to surface density variations, this feature might be related to the edge of a cavity-like structure around A1. Such cavities are expected to form in binary systems, and can have complex shapes depending on the system's dynamical properties \citep{2019MostaGas,2019MatsumotoStructure}. Alternatively, the reduced optical depth regions around the A1 circumstellar disk could also correspond to a reduction in the overall $\kappa$ due to the higher temperatures leading to ice mantle evaporation in these spots \citep{2003Semenov,2016CiezaImagingWaterSnow}. This scenario would also be consistent with the high dust temperatures derived in this work, which can reach values over 100 K in the regions with reduced optical depth around p1 and p2 (Figures~\ref{fig:tau_temp_beta_ism} and~\ref{fig:temp_tau_1.3}). \\

Assuming a gas to dust ratio of 100 and a $\kappa_{\rm 223 GHz}=0.89$ cm$^2$ g$^{-1}$ \citep{1994OssenkopfDust}, the total gas mass in the circumbinary material derived from this analysis is $\sim$0.06-0.13 M$_{\odot}$. However, these values can be reduced by a factor of several and up to an order of magnitude by assuming different values for $\kappa_{\rm 223 GHz}$\footnote{Due to some level of grain growth with respect to grains in the ISM or different dust compositions for example (e.g., \citealt{2019AgurtoGangasRevealing,2019ValdiviaIndirect,2021ZamponiYoung}).}.\\

\subsubsection{Temperature and COM emission}
\label{sec:mol_fit}

\cite{2020MaureiraOrbital} presented integrated intensity and velocity maps for 3 mm transitions of the complex organics HNCO, NH$_2$CHO and t-HCOOH at a resolution of 13 au. Here we re-examine how the intensity distribution of molecular lines relates to the dust temperature enhancements and optical depth variations. We performed a Gaussian fit to the molecular line cubes presented in  \cite{2020MaureiraOrbital} considering only the pixels for which $S/N\geqslant3$. After the fit, we also masked out pixels for which the derived values for the free parameters (peak intensity and velocity dispersion) was less than five times their corresponding errors from the fit. The bottom panels of Figure~\ref{fig:tau_temp_beta_ism} show the resultant line intensity peak distribution for HNCO(5$_{3, 2}$-4$_{3,1}$) and NH$_2$CHO(5$_{1, 4}$-4$_{1,3}$). Figure~\ref{fig:spec_fit_mols} in the Appedix shows the spectra and resultant fit averaged over a beam at the location of the continuum peaks p1, p2 and p3.\\

The dust hot spots related to p1 and p2 correspond well with enhancements in the line intensity from these species, particularly for p2. This is consistent with the scenario that these COMs are evaporated from the icy mantles on the surface of dust grains in the hot spots. In the case of p3, for which lower dust temperatures are inferred, there is no related COM emission which could be due to actual lower temperatures therein or the sensitivity ($\sim$ 15 K) of the molecular line observations \citep{2020MaureiraOrbital}.  \\

\section{Discussion}
\label{sec:discussion}

\subsection{The origin of the hot dust spots}

It is clear that the observations are not consistent with a smoothly decaying temperature profile expected in the standard scenarios of radiative heating and steady-state viscous heating. In the following, we discuss other scenarios involving mechanical heating, namely shocks and dissipation following disk fragmentation. 

\subsubsection{Are the hot spots consistent with shocks?}
\label{sec:theo_shocks}

Shocks can generate large local temperature enhancements, and are thus a likely culprit of the observed hot spots. Moreover, shocks are prevalent in numerical simulations of binary systems, resulting from gravitational interactions between the binary stars and the surrounding gas material. Shocks occur in the forms of steepened spiral density waves in circiumstellar disks \citep{2016Ju}, accretion streams onto the circumstellar disks \citep{2019MostaGas}, and impact streams close to the inner edge of the circumbinary disk \citep{2012Shi}. However, most of the simulations to-date adopt an isothermal or a locally isothermal equation of state, unable to provide realistic shock temperatures that can be directly compared with our observations. Below we give a simple estimation of the shock temperature and structure for our system based on theoretical expectations.\\

We expect the shock speed to be $v_s \approx v_k \sin(\theta)$, where $v_k$ is the Keplerian speed, and $\theta$ is the pitch angle of the shock. Assuming the hot spots to be at a deprojected distance of 80 AU from the protostars (similar to the binary separation), a combined protostellar mass of 4 $M_\odot$ \citep{2020MaureiraOrbital}, and a pitch angle of $30^{\circ}$, this gives a shock speed of $v_s \approx 3~\mathrm{km/s}$. This shock speed is comparable to the FWHM of the molecular lines ($\sim3-5~\mathrm{km/s}$) at the position of the hot dust spots measured from a Gaussian fit (Section~\ref{sec:temp_tau_derivation}). The shock speed is much higher than the sound speed in the (pre-shocked) disk, at about $0.6~\mathrm{km/s}$ assuming a gas temperature of 100 K. In this case, the mach number ${\cal M} \gg 1$, and the post-shock temperature in a strong adiabatic shock can be approximated by \citep[e.g.][]{2011Draine},
\begin{equation}\label{eq:T_s}
    T_s \approx \frac{2(\gamma-1)}{(\gamma+1)^2} \frac{\mu v_s^2}{k_{\rm b}} \approx 400~\mathrm{K},
\end{equation}
where $\gamma = 7/5$ is the adiabatic index for molecular hydrogen, $\mu = 2.3 m_\mathrm{H}$ is the mean molecular weight accounting for helium, $m_\mathrm{H}$ is the mass of hydrogen nuclei, and $k_{\rm b}$ is the Boltzmann constant. We note that  the shock temperature is very sensitive to the pitch angle: the maximum post-shock temperature of about $1.8\times 10^3~\mathrm{K}$ can be reached for a perpendicular shock with $\theta=90^{\circ}$.

In reality, $T_s$ is the gas temperature immediately behind the shock front. Away from the shock front, the post-shock gas cools down over time. Taking a gas surface density of $\Sigma_\mathrm{gas} \approx 100~\mathrm{g\ cm^{-2}}$ (see Section \ref{sec:alpha_var}), and Rosseland mean opacity of $\kappa_{\rm R} \approx 4.5~\mathrm{cm^2\ g^{-1}}$ \cite[][
for gas temperature $100~\mathrm{K} \lesssim T_\mathrm{gas} \lesssim 2000~\mathrm{K}$]{2003Semenov}, the optical depth is $\tau \approx 450$. In the optically thick regime, the cooling time \citep{2009Rafikov},
\begin{equation}\label{eq:t_cool}
    t_\mathrm{cool} = \frac{\Sigma_\mathrm{gas} c_s^2}{\sigma T_s^4} \tau \approx 15~\mathrm{yr}.
\end{equation}
This gives a shock width of $L_s \approx t_\mathrm{cool}c_s \approx 5 ~\mathrm{AU}$, where $c_s$ is the sound speed at the post-shock temperature $T_s$. It is worth noting that both the cooling time and the shock width are proportional to the dust opacity, which depends on the size distribution and composition of dust grains.\\

In this very simple estimation, the shock temperature and width are broadly consistent with the observational constraints. It is important to note that one should also expect to see an enhancement of density due to shocks. In our observations, we do not find strong variations in $\alpha$ (or $\tau$) at the position of the temperature peaks (Figures \ref{fig:sourceA_alpha_kde} and \ref{fig:tau_temp_beta_ism}). In strong shocks, the ratio between the pre-shock and post-shock temperature is proportional to the square of the shock velocity (Equation (\ref{eq:T_s})), while the density ratio is capped at a maximum of $(\gamma + 1)/(\gamma - 1) = 6$. Depending on the shock velocity and pre-shock conditions, it is possible that the temperature enhancement is two or more times higher than the corresponding density enhancement. Moreover, the opacity is expected to drop by a factor of $\sim 5$ when the temperature increases from $\sim 150$ K to $\sim 500$ K, due to the evaporation of ice mantle and volatile organics  on dust grains \citep{2003Semenov}. This opacity drop can compensate for the increase of density (and column density) due to shocks, resulting in no significant increase in the optical depth $\tau$.  \\

We note that large scale infall, such as the accretion flow or streamer observed in \citet{2022MurilloColdaccretionFlow}, may also result in shocks when material lands on the circumbinary structure or the edge of a circumstellar disk. Such scenario, unlike the gravitational interactions between the binary and the gas discussed above, applies also to single sources. In binary systems however, the shock structure is likely more complex and dependent on the dynamical properties of the system. We note that the analysis on the dependence of shock temperature on Mach number discussed above would also apply in shocks due to accretion streamers and the shock velocity would also be comparable to the Keplerian velocity.\\

In shocks, the heating acts directly on the gas, and dust is heated through collisions with the gas. The cooling, on the other hand, is mainly through dust thermal emission. In this case, we expect the gas temperature $T_g$ to be larger than the dust temperature $T_d$. In steady state, the dust cooling rate $\Lambda_d$ equals the dust heating rate by gas $\Phi_{gd}$. Since $\Lambda_d\propto 1/t_\mathrm{cool}\propto 1/\tau$ (Equation \ref{eq:t_cool}) and $\Phi_{gd} \propto n_g^2 (T_g - T_d)$ where $n_g$ is the gas number density \citep{2001Goldsmith}, we expect the dust temperature to closely follow the gas temperature in our observed regions, where the gas density is high and optical depth is also high (similar to the assumptions used in self-gravitating disks in \citealt{2011IleeChemistry}).\\

In general, the shock structure is very sensitive to several uncertain parameters, such as the shock velocity (which depends on the binary masses), pitch angle, and dust opacity. We plan to perform numerical simulations and radiation transfer calculations in the future to model the shocks and compare with observations in detail. This will help us understand under which conditions shocks can produce the observed gas and dust hot spots. \\

\subsubsection{Can the hot spots be created by gravitational instability?}

Gravitational instability can lead to fragmentation in young massive disks, resulting in the formation of clumps that may eventually form a stellar companion or a planet \citep[see review by][and references therein]{2016KratterLodato}. In a collapsing clump, the gravitational energy released may heat up the gas, causing local temperature enhancements \citep{2012Zhu}. Can the hot spots we observed be generated by the gravitational instability? We offer some theoretical estimates below to argue that this is an unlikely scenario.\\

Assuming the circumbinary material is a rotating disk (suggested by the flattened geometry and rotation observed in \citealt{2020MaureiraOrbital}), the standard parameter for quantifying the degree to which the disk is self-gravitating is \citep{1964Toomre},
\begin{equation}
    Q = \frac{c_s \Omega}{\pi G \Sigma_\mathrm{gas}},
\end{equation}
where $\Omega$ is the epicyclic frequency, which equals the angular frequency in Keplerian disks. Gravitational instability in general requires $Q < 1$. In order for the disk to fragment and create gravitationally bound objects, an additional criterion requires the cooling time to be relatively short compared to the orbital time, expressed $\beta = \Omega t_\mathrm{cool} \lesssim 3$ \citep{2017Baehr, 2017Forgan}. Assuming a background disk temperature of 100 K \citep{2020OyaSubstructures}, the hot spots to be at a deprojected distance of 80 AU from the protostars (similar to the binary separation), a total mass of the binary of $4~M_\odot$ \citep{2020MaureiraOrbital}, $\Sigma_\mathrm{gas} = 100~\mathrm{g/cm^{-2}}$ (see Section \ref{sec:alpha_var}), and optical depth $\tau=450$ (see Section \ref{sec:theo_shocks}), we obtain $Q=1.6$ and $\Omega t_\mathrm{cool}=16$. Given that the disk surface density we choose is already an upper limit assuming the ISM $\beta=1.7$ and gas-to-dust ratio of 100, it is unlikely that the circumbinary disk is gravitationally unstable and undergoing fragmentation. Spiral arm features may form in disks with $Q=1.6$ \citep{2010Kratter}. However, with the long cooling time, the spiral arms are expected to be tightly wound and only have almost sonic Mach numbers \citep{2009Crossins}, unable to produce the prominent local temperature enhancements we observe.\\

If the hot spots are clumps collapsing under self-gravity, it requires the clumps to be gravitationally bound against both the thermal pressure support and the tidal disruption. Gravitational collapse against thermal pressure requires $G M_{cl}^2/R_{cl} > c_s^2 M_{cl}$, where $M_{cl}$ and $R_{cl}$ are the clump mass and radius. This requires the clump surface density $\Sigma_{cl} > c_s^2/G R_{cl}$, where $\Sigma_{cl}=M_{cl}/R_{cl}^2$. Using a conservative estimate of the clump temperature at 200 K \citep{2020OyaSubstructures} and $R_{cl}=10~\mathrm{AU}$, this gives $\Sigma_{cl} > 700~\mathrm{g/cm^{-2}}$. The stability against tidal disruption requires $R_{cl} < R_H = (M_{cl}/3 M_*)^{1/3} a$, where $R_H$ is the Hill radius and $a$ is the semi-major axis. This gives $\Sigma_{cl} > 3R_{cl}M_*/a^3 = 2100 ~\mathrm{g/cm^{-2}}$, using $M_*=4~M_\odot$ as the total binary mass and $a=80~\mathrm{AU}$. Such high surface density enhancement of more than an order of magnitude is unlikely since no significant increase in optical depth is observed at the location of the hot spots. Thus, we prefer the scenario where the hot spots are created by shocks instead of gravitational instability.

\subsection{Mechanical heating in Hot Corinos}

%Mechanical heating in embedded and actively accreting protostars}
\label{sec:mech_heating_hotcorino}
%\subsubsection{Shocks and COM emission}

IRAS 16293 A1-A2 is well-known for its rich `hot corino' chemistry exhibiting many lines from COMs from which temperatures over 100 K are derived \citep{2003CazauxHotCore,2004BottinelliNearArcsecond,2005ChandlerIRAS16293,2012PinedaFirst,2016JorgensenPils}. The presence of these species in the gas-phase in this as well as other hot corinos is thought to be due to the sublimation of ices on dust grains, resulting in the desorption of COMs, or due to the subsequent hot gas-phase chemistry. Both radiative heating and mechanical heating can act to heat up the dust grains \citep{2017MiuraComprehensive,2018JacobsenALMAPILS}.  \\

It has been difficult to unambiguously disentangle the heating mechanism in hot corinos because most observations do not resolve the compact COM emitting region at disk scales \citep{2019JacobsenOrganic,2020Belloche2020Questioning,2020BianchiFAUSTI,2021MartibDomenechHotCorino,2021YangPeaches,2022VastelHotMethanol}. Although some observations have been able to identify COM and sulfur-bearing species arising preferentially in the transition between the disk and the infalling envelope material in which shocks are expected \citep{2014NaturSakai}, evidence of elevated dust temperature is missing. Our observations show direct evidence of hot dust spots which are consistent with shocks and correlated with hot gas spots derived from H$_2$CS lines in \cite{2020OyaSubstructures}. Shock heating can thus provide a natural explanation of the COM emission peaks observed outside the circumstellar disks\footnote{We note that while we do not see COM emission coming from within the disks (e.g., Figure~\ref{fig:tau_temp_beta_ism}), their presence there cannot be ruled out as the disks are optically thick which can hide the molecular emission \citep{2019SahuImplications,2020DeSimoneHotCorinos}.}.\\

%as well as other single and multiple protostellar systems \citep{2020PiendaPeremb2streamer,2022Valdivia-MenaPeremb50streamer,2022PinedaPPVII}

This Class 0 binary is not the only known hot corino source in which high-resolution observations have suggested the presence of mechanical heating. \cite{2022VastelHotMethanol} argued the hot methanol resolved down to 50 au around the Class I binary BHB2007-11 (separation of 28 au) is produced by shocks. In the case of the Class 0/I SVS 13 A system (separation of 90 au), some COMs have been observed to emit in the circumstellar disks but also in a more extended region associated with the circumbinary gas \citep{2022DiazRodriguezPhysical}. Such an extended component could be, as in IRAS 16293 A, associated with shocks \citep{2022BianchiSVS13}. We note that both shocks and accretion heating could be further enhanced by accretion `streamers' which have been detected towards IRAS 16293-2422 and SVS 13 A so far \citep{2022MurilloColdaccretionFlow,2022DiazRodriguezPhysical}. For instance, the Class I binary L1551 IRS 5 (separation of 54 au, \citealt{2019CruzL1551IRS5}) is an FUor-like object \citep{2018AConnelleyFUOrlike} with measured high accretion rates ($2-6\times10^{-6}$ M$_{\odot}$yr$^{-1}$) toward each component \citep{2005LiseauL1551IRS5}. This can lead to elevated temperatures in the circumstellar disks due to viscous heating (e.g, \citealt{2021LiuFuOri}) and could also explain the compact COM emission observed towards each disk \citep{2020BianchiFAUSTI}. These sources discussed so far correspond to all reported hot corinos which have been resolved into close separation ($<100$ au) binaries \citep{2021MartinDomenechHotCorino2021}. \\

Close separation systems such as IRAS 16293 A, especially the more embedded ones, would be more likely to produce the type of shocks discussed in Section~\ref{sec:theo_shocks}. Suggestive of this, we note that 5 out of the 7 multiple Class 0 systems with separations below 100 au in the Perseus VANDAM survey \citep{2016TobinVLA}, were also revealed to harbor emission from CH$_3$OH (Per-emb 2) and in some cases from CH$_3$CN or CH$_3$OCHO as well (Per-emb 18, Per-emb 17, Per-emb 5 and SVS13A). The COM emission in those observations (except for SVS13A) remains unresolved with a resolution of $\sim$200 au \citep{2021YangPeaches}. Higher-resolution observations are required to assess the origin of these lines. For the remaining two cases (L1448 IRS3C and L1448 IRS3B), possible reasons that could lead to non-detection of any COM are that the regions remain too optically thick (e.g., \citealt{2020DeSimoneHotCorinos}) or that the intensities fall below current sensitivities. In addition, it could be that the physical conditions do not result in strong shocks, that the surrounding gas densities are below what is required to heat the dust mantles in the grains through collisions, or due to time variability of the shocks. Nevertheless, it seems possible that shocks are indeed present in all 7 sources given that in all of them emission from sulfur-bearing species such as SO were also detected \citep{2021YangPeaches}.\\

For some of the remaining hot corinos which are single sources or in multiples at wider ($>100$ au) separations there is also evidence of mechanical heating. The high gas and dust temperatures in IRAS 16293 B, located $\sim$700 au to the northwest of A1-A2 (Figure~\ref{fig:obs_1mm_3mm}), have been successfully reproduced considering a massive (0.3 M$_{\odot}$), optically thick and gravitationally unstable disk \citep{2021ZamponiYoung}. The high temperatures in the disk are due to accretion heating and shocks. Similarly, a hot gravitationally unstable disk was also proposed to explain the dust emission towards the Class 0 hot corino HH212 \citep{2021LinInferring}. Finally, a heating mechanism associated with the disk was also tentatively suggested for the case of NGC 1333 IRAS4A2 and NGC 1333 IRAS 4B in \cite{2020Belloche2020Questioning}.\\

\subsection{Mechanical heating during the embedded protostellar stages}
\label{sec:mech_heat_protostars}

Given that non-radiative heating in the form of shocks is not only expected in binary interactions but also due to envelope/disk and spiral arms/disk interactions, \citep{2011IleeChemistry,2017MiuraComprehensive,2015EvansGravitational} it might be generally important in the modelling of the inner regions in both multiple and single protostellar sources. This is supported by the increasing number of studies suggesting a shock origin for the unresolved emission from COMs and sulfur-bearing species showing complex line profiles and elevated gas temperatures in both Class 0 and I sources (\citealt{2019ArturdelaVillarmoisPhysical,2019OyaSulfur,2022VastelHotMethanol,2022DiazRodriguezPhysical,2022Valdivia-MenaPeremb50streamer,2022HsiehPRODIGESVS13A}). Moreover, in an unbiased chemical survey of 50 embedded protostars in Perseus \citep{2021YangPeaches}, 58\% show emission from COMs. Similarly, sulfur-bearing species such as SO and SO$_2$, predicted to be shock tracers \citep{2017MiuraComprehensive,2021vanGelderModelingShocks}, are detected in $\sim80\%$ of the protostars in the survey. %Future high-resolution observations towards more sources will tell if these lines are indeed tracing shocks or if they can be modelled with accretion heating. 
We note that radiative heating can also produce high temperature and COM emission in disks, for example  during outburst events when the luminosity becomes 10-100$\times$ higher (e.g., \citealt{2017FrimannProtostellar}). Nonetheless, in the case of radiative heating as well as of viscous heating (e.g., if the outburst is due to high accretion rates at disk scales), we expect the temperature to smoothly decrease away from the protostar(s), instead of producing localized hot spots \citep{2015HarsonoVolatile,2017RabTheChemistry,2017FrimannProtostellar}.\\

Theoretical works and numerical simulations also show that dissipation processes during the formation of the disk and the high-accretion rates can also contribute to heating up young embedded disks \citep{2007DullemondModels,2015HarsonoVolatile,2021XuFormation,2021ZamponiYoung}. As previously discussed in Section~\ref{sec:mech_heating_hotcorino}, such scenario successfully reproduces the observations for the hot disks towards the Class 0 IRAS 16293 B \citep{2021ZamponiYoung} and HH212 \citep{2021LinInferring} disks. \\

Understanding whether mechanical heating in the form of shocks, viscous dissipation, adiabiatic compression, etc., is prevalent during the early stages is critical for the derivation of physical properties of disks, such as temperature, density, mass and grain size \citep{2018SeguraCoxVandamV,2019MauryCharacterizing,2019AgurtoGangasRevealing,2020TobinVandamOrion,2022SheehanVLAALMA}. Taking into account the accretion heating in gravitationally unstable disks can lead to an excellent match of the observed fluxes at mm wavelengths and it results in higher masses (in a factor of $\sim$ 10) than those derived using passively heated disks models for the same set of observations. This can lead to opposite conclusions regarding the gravitational stability of the disks \citep{2022SheehanVLAALMA,2022XuTesting}. As pointed out by \cite{2022XuTesting}, while the fully radiative and fully mechanical heating models represent two extremes and both mechanisms might be important in reality, the significant difference in the conclusions highlights the need to understand their roles better. Future multi-wavelength observations are clearly needed to provide direct evidence of the prevalence of mechanical heating at the early stages of star formation and to break degeneracies arising from unresolved structures. \\

Another important example is the interpretation of low spectral index values. Low values, even below the optically thick limit of 2, can be naturally explained by a disk undergoing mechanical heating \citep{2021LinInferring,2021ZamponiYoung,2021XuFormation}, without requiring any grain growth or scattering effects \citep{2017LiSpectralIndex,2018GalvanMadridEffects,2021ZamponiYoung}. These disks are usually optically thick and massive, contrary to the common optically thin assumption used to calculate masses from the dust emission. \\

\subsection{Grain sizes during protostellar stage}

High-resolution dust continuum observations at two or more wavelengths towards Class II disks have shown that they already contain a significant amount of grains with mm up to cm sizes. Direct evidence of this comes from observations in which the disks are well resolved. Those studies reveal a radial dependence on $\beta$, showing values of $\beta$ of $\lesssim1$ close to the center and up to $\sim$2 on the outer edge \citep{2012PerezConstraints,2019CarrascoRadial,2021MaciasCharacterization,2021TazzariMultiwavelength}. \\

In Class 0 and I sources, several studies attempt to constrain the dust properties at envelope scales using observations with resolutions ranging from 3000 au to 200 au \citep{2009KwonGrainGrowth,2011AShirleyMustang,2014MiotelloGraingrowth,2019AgurtoGangasRevealing,2019GalametzLowdust}. These studies apply radiation transfer and {\it uv}-space analysis techniques to separate envelope and disk structures. In some cases $\beta$ values below 1 at 200 to 2,000 au scales \citep{2019GalametzLowdust,2014MiotelloGraingrowth} are inferred. In others, the inferred value for $\beta$ is consistent with the ISM value of $\sim$1.7 \citep{2019AgurtoGangasRevealing}. While the former suggests grain growth up to millimeter sizes in envelopes as early as the Class 0 stage, the latter conclude no grain growth or that only grains with sizes up to few 100 $\mu$m are present. However, most of these observations do not resolve envelope from disk(s) and assumptions had to be made to account for embedded unresolved structures and their physical properties. As discussed in Section~\ref{sec:mech_heat_protostars}, the differences in the model assumptions can greatly affect the grain sizes \citep{2017LiSpectralIndex}.   \\

For our particular source, if $\beta$ is constant in the circumbinary material we derived a lower limit to $\beta$ of 1.3 (Section~\ref{sec:analysis}). Alternatively, if $\beta$ is changing throughout the circumbinary material and considering the measured median for $\alpha$ of 3.1, some of the regions might have $\beta\sim1$ or slightly below. In either case these values are in agreement with some of the previous results towards other Class 0 and I sources \citep{2009KwonGrainGrowth,2019AgurtoGangasRevealing}. Our results for the circumbinary material are also consistent with recent work using similarly high-resolution ALMA observations towards Class 0 disks for which $\beta$ values between 1 and 1.6 are inferred \citep{2021LinInferring,2021ZamponiYoung,2022OhashiFormation}. Likewise, similar results were recently obtained for at least one Class I disk \citep{2022OhashiNoEvidence}. On the other hand, the value for $\beta$ measured here is significantly higher than the values inferred recently for 10 Class 0 sources; most of them showing $\beta=0.5$ or below at scales of several 100 au \citep{2019GalametzLowdust}.  \\

Whether the $\beta$ values inferred here could be related to the presence of millimeter grain sizes depends strongly on the dust properties and distribution of sizes. Generally, $\beta$ values firmly below 1 are a more definitive indication of mm/cm grain sizes \citep{2014TestiDustEvolution}. If mm/cm grain sizes are not present, grain sizes up to few 100 $\mu$m would also be in agreement with our measured $\beta$ \citep{2019AgurtoGangasRevealing}. This last scenario, not requiring mm/cm grain sizes, would also be consistent with recent numerical simulations that follow the dust evolution during the collapse of a prestellar core down to disk densities  \citep{2022BateDustCoagulation} as well as recent analytical considerations \citep{2022SilsbeeDust}. However, given the high densities inferred in the circumbinary material, such studies do not fully eliminate the possibility of grain growth to mm sizes. \\

Another scenario that could lead to a lack of mm grain sizes near the circumstellar disks, despite the high densities, is the evaporation of the dust ice mantles. The mechanism discussed in the literature is the reduction of the velocity limit for fragmentation in ice-free grains, leading to shattering and subsequent replenishment of small grains \citep{2015BanzattiDirect,2016CiezaImagingWaterSnow}. 
This effect can be observed as an increase of $\alpha$ (or $\beta$). For our source, slightly higher values of $\alpha$ are indeed observed around A1 (Figure~\ref{fig:1mm_3mm_alpha}). The main constituents of ice mantles, CO$_2$ and water, require high dust temperatures ($>70-150$ K) to evaporate \citep{2001FraserThermaldesorption,2011FayolleLaboratory,2014MartinDomenechThermaldesorption,2018PotapovTemperature}. These high temperatures are in better agreement with the derived temperatures in the case of $\beta=1.3$ (Figure~\ref{fig:temp_tau_1.3}). However, as pointed out in \cite{2021LiuFuOri}, recent laboratory results suggest that bare grains might be at least as prompt to stick together and grow as those covered by ice \citep{2018GundlachTensile,2019MusiolikContacts,2019SteinpilzSticking,2021PillichDrifting}. Suggestive of this, \cite{2021LiuFuOri} derived grain sizes $>1.6$ mm in the inner disk around Fu Ori, with dust temperatures around  400 K. Future sensitive observations at longer wavelengths, for example, with ALMA band 1, can improve our constraints on the maximum grain sizes present in the derived hot spots towards IRAS 16293 A, helping to address further the question of grain growth in regions with evaporated ice mantles.

%I think it would be useful here again to mention that the evaporation of ice mantles can also lead to the reduction of grain sizes, depending on how thick the ice is. Which could also be an explanation for the lack of the lack of mm-sized particles. It's interesting that the 'ridge' in Tau shown in Figure 4 seems to approximately follow the 50 K temperature contour. At the risk of over interpreting the results, if there was an outburst that is now fading, that temperature contour could have previously been the 80 or 100 K contour at which point CO2 and water, respectively, would be evaporated, which constitute most of the ice. Or if beta is a bit lower, then the temperature at that same contour could be higher as well and the same scenario could be at play.

\section{Summary and conclusions}
\label{sec:conclusions}

In this work we analyzed ALMA observations at 1.3 mm (12.5 au resolution) and 3 mm (6.5 au resolution) towards the close binary A1-A2 within the triple Class 0 system IRAS 16293-2422. Continuum substructures in the circumbinary material around the compact disks correspond to localized dust hot dust with individual temperatures of at least 122, 87 and 49 K. Depending on the assumed dust opacity index, these values can increase by a factor of several. The location of the dust hot spots matches high gas temperature peaks and compact COM line emission previously reported in the literature. This present evidence is consistent with theoretical predictions from shock heating instead of the commonly assumed irradiation heating. Future work comparing simulations with observations will help set tighter constraints on the role of shocks in heating the dust and gas and the dependence with the properties of the system. In addition, the analysis of the spectral index map indicates dust opacity index $\beta$ values around one or higher with no significant spatial variations thus providing no firm indication of millimeter-sized grains.\\

This work, as well as other recent studies using high-resolution multi wavelength continuum observations show that mechanical heating in the form of shocks and dissipation powered by accretion might be significant at the early stages of protostar and disk formation in both multiple and single systems. Ignoring the contribution of mechanical heating can lead to significantly different estimates of disk temperatures, and thus also other physical quantities such as typical optical depths, dust grain sizes and disk masses. 
Future multi-wavelength observations are needed to understand the prevalence of mechanical heating at the early stages of star formation with implications for the chemical and physical properties measured at disk scales during the early stages. 
\\

\section{acknowledgments}
M.M, J.E.P., P.C., M.G., K.S. and D. S.-C. acknowledge the support by the Max Planck Society. H.B.L. is supported by the Ministry of Science and
Technology (MoST) of Taiwan (Grant Nos. 108-2112-M-001-002-MY3 and 110-2112-M-001-069-). M.G. thank Kaitlin Kratter and Lucio Mayer for discussions at the KITP binary22 program. This research was supported in part by the National Science Foundation under Grant No. NSF PHY-1748958.

This paper makes use of the following ALMA data: ADS/JAO.ALMA\#2017.1.01247.S and ADS/JAO.ALMA\#2016.1.00457.S. ALMA is a partnership of ESO (representing its member states), NSF (USA) and NINS (Japan), together with NRC (Canada), MOST and ASIAA (Taiwan), and KASI (Republic of Korea), in cooperation with the Republic of Chile. The Joint ALMA Observatory is operated by ESO, AUI/NRAO and NAOJ. 


\vspace{5mm}
\facilities{ALMA}


\software{aplpy \citep{aplpy2012,aplpy2019},
          astropy \citep{astropy:2013,astropy:2018},  
          CASA \citep{2007McMullinCASA}}



\appendix
\restartappendixnumbering

\section{Spectral index distribution}

\begin{figure}
   \centering
     \includegraphics[width=0.4\textwidth]{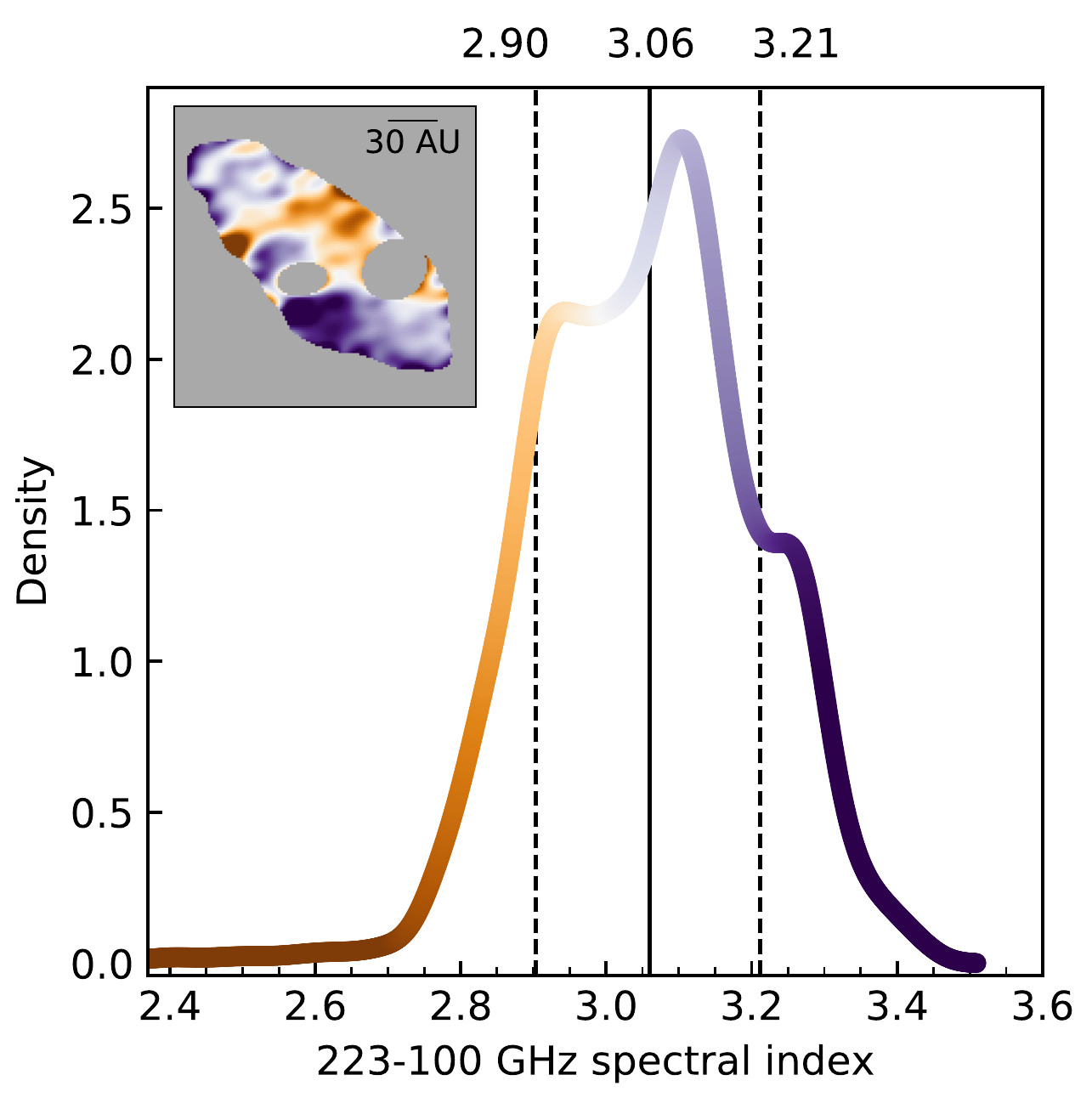}
      \caption{Spectral Index distribution of the circumbinary material obtained through a Kernel Density Estimation (KDE). The inset in the top left
corner shows the spectral index map with the pixels used to calculate the KDE colored according to main panel x-axis. The median and values enclosing 68\% of the values around the median are shown in the top and marked with solid and dashed lines, respectively.}
         \label{fig:sourceA_alpha_kde}
\end{figure}

Figure ~\ref{fig:sourceA_alpha_kde} shows the spectral index $\alpha$ distribution of the circumbinary material obtained through a Kernel Density Estimation (KDE).

\section{Temperature and optical depth map with $\beta=1.3$}

\begin{figure*}
   \centering
     \includegraphics[width=1\textwidth]{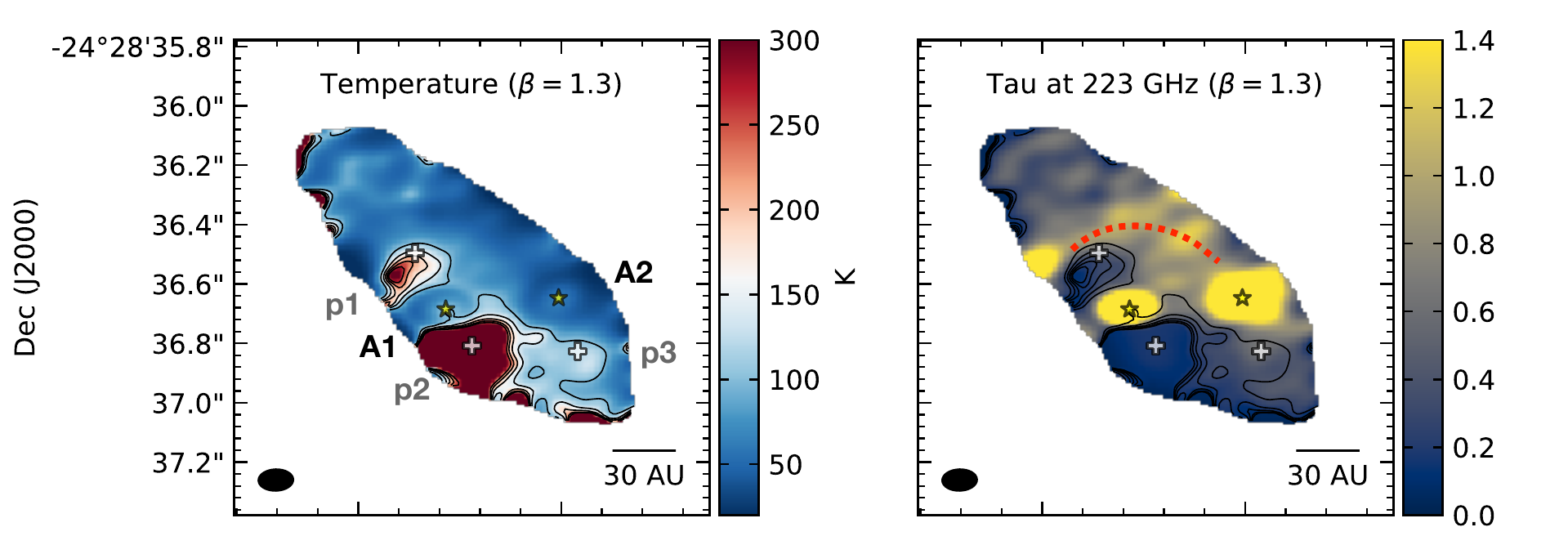}
      \caption{Similar to Figure~\ref{fig:tau_temp_beta_ism}, the panels correspond to the derived dust temperature and optical depth at 223 GHz maps assuming a constant dust opacity index ($\beta$) of 1.3. Black contours correspond to dust temperature of 100, 140, 180, 220 and 260 K. The red dotted line highlights the region where the optical depth values at 1.3mm are the highest outside the circumstellar disks. The beam for each map is drawn in the bottom left corner.}
        \label{fig:temp_tau_1.3}
\end{figure*}

Figure~\ref{fig:temp_tau_1.3} shows the derived dust temperature and optical depth at 223 GHz assuming a constant dust opacity index $\beta=1.3$. This $\beta$ value corresponds to the minimum value that is able to match the observed spectral index distribution when assuming constant $\beta$. Thus, the absolute dust temperature and optical depths at 223 GHz correspond to an upper and lower limit, respectively. Black contours in all panels follow the dust temperature (left panel) and are drawn at 100, 140, 180, 220 and 260 K, respectively. The star symbols mark the position
of the protostars A1 and A2.

\section{Spectra and Gaussian fit at continuum peaks}

\begin{figure*}
   \centering
     \includegraphics[width=1\textwidth]{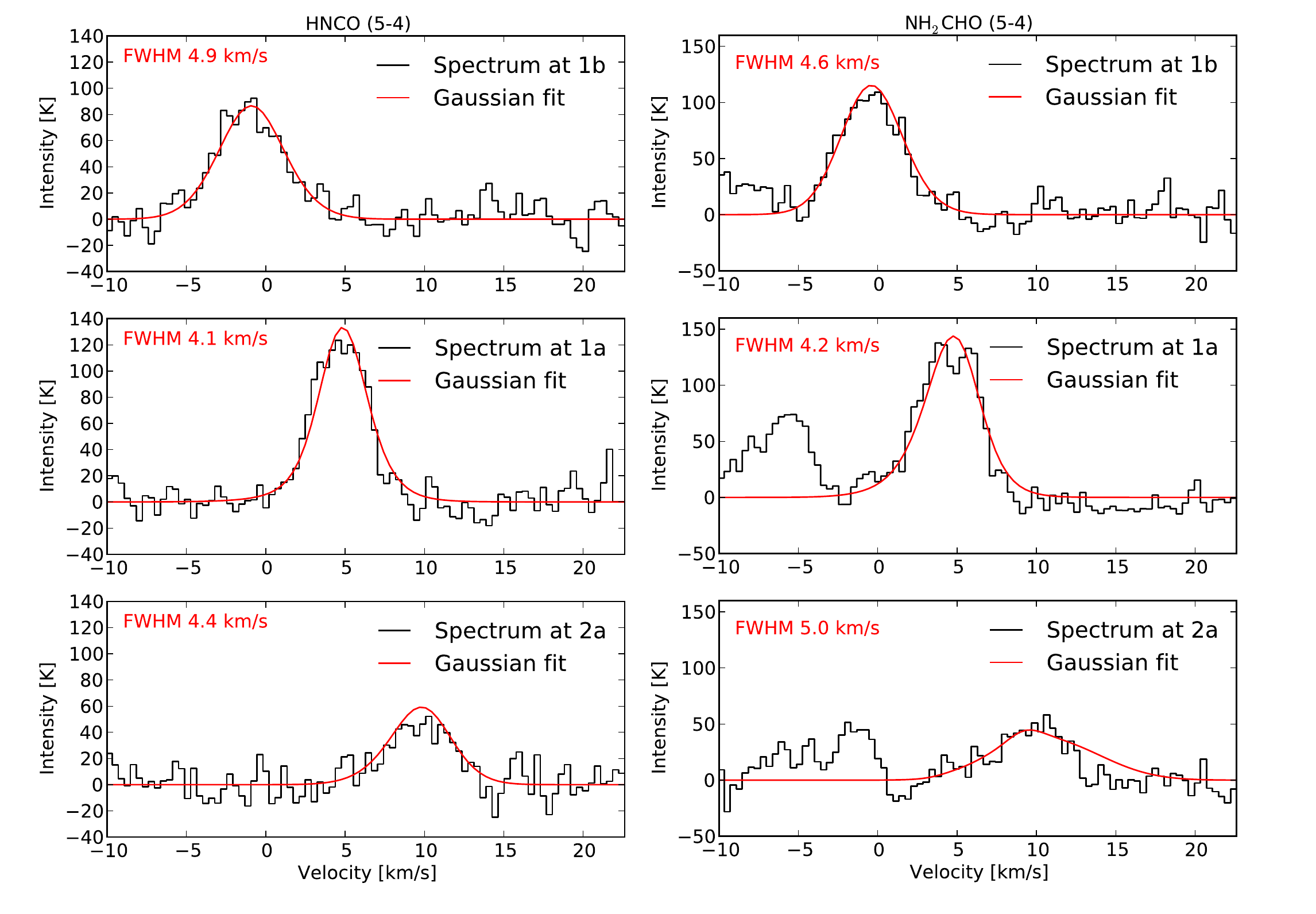}
      \caption{Beam averaged spectra for HNCO(5$_{3, 2}$-4$_{3,1}$) and NH$_2$CHO(5$_{1, 4}$-4$_{1,3}$) at the continuum peaks p1, p2, and p3 (see Figure~\ref{fig:obs_1mm_3mm}). The results from the Gaussian fit discussed in Section~\ref{sec:mol_fit} is shown with a red line. The resultant FWHM from the fit is shown in the upper left corner in each panel.}
        \label{fig:spec_fit_mols}
\end{figure*}

Figure~\ref{fig:spec_fit_mols} shows the beam averaged spectra for HNCO(5$_{3, 2}$-4$_{3,1}$) and NH$_2$CHO(5$_{1, 4}$-4$_{1,3}$) at the location of the continuum peaks p1, p2, and p3 marked in Figure~\ref{fig:obs_1mm_3mm}. The results from the Gaussian fit discussed in Section~\ref{sec:mol_fit} are shown with a red line. The resultant FWHM from the fit is shown in the upper left corner in each panel.

% \section{Using Chinese, Japanese, and Korean characters}

% Authors have the option to include names in Chinese, Japanese, or Korean (CJK) 
% characters in addition to the English name. The names will be displayed 
% in parentheses after the English name. The way to do this in AASTeX is to 
% use the CJK package available at \url{https://ctan.org/pkg/cjk?lang=en}.
% Further details on how to implement this and solutions for common problems,
% please go to \url{https://journals.aas.org/nonroman/}.


\bibliography{references}{}
\bibliographystyle{aasjournal}



\end{document}